\documentclass[prl,aps,preprint,superscriptaddress]{revtex4-2}

\usepackage{amsmath}
\usepackage{amssymb} 
\usepackage{color}
\usepackage{makeidx}
\usepackage{amsfonts}
\usepackage{bmpsize}
\usepackage{subfigure}
\usepackage{epsfig}

\usepackage[colorlinks,hyperindex]{hyperref}
\usepackage[squaren,Gray]{SIunits}
\usepackage{graphicx}
\usepackage{float}

\usepackage[normalem]{ ulem }
\usepackage{soul}

\date{\today}

\begin{document}

\title{Specific Heat of Thin Phonon Cavities at Low Temperature: Very High Values Revealed by ZeptoJoule Calorimetry}

\author{Adib Tavakoli}
\affiliation{Institut N\'EEL, CNRS, 25 avenue des Martyrs, F-38042 Grenoble, France}
\affiliation{Univ. Grenoble Alpes, Inst NEEL, F-38042 Grenoble, France}

\author{Kunal J. Lulla}
\affiliation{Institut N\'EEL, CNRS, 25 avenue des Martyrs, F-38042 Grenoble, France}
\affiliation{Univ. Grenoble Alpes, Inst NEEL, F-38042 Grenoble, France}

\author{Tuomas Puurtinen}
\affiliation{Nanoscience Center, Department of Physics, University of Jyv\"{a}skyl\"{a}, PO Box 35, FI-40014 Finland}

\author{Ilari Maasilta}
\affiliation{Nanoscience Center, Department of Physics, University of Jyv\"{a}skyl\"{a}, PO Box 35, FI-40014 Finland}

\author{Eddy Collin}
\affiliation{Institut N\'EEL, CNRS, 25 avenue des Martyrs, F-38042 Grenoble, France}
\affiliation{Univ. Grenoble Alpes, Inst NEEL, F-38042 Grenoble, France}

\author{Laurent Saminadayar}
\affiliation{Institut N\'EEL, CNRS, 25 avenue des Martyrs, F-38042 Grenoble, France}
\affiliation{Univ. Grenoble Alpes, Inst NEEL, F-38042 Grenoble, France}

\author{Olivier Bourgeois}
\affiliation{Institut N\'EEL, CNRS, 25 avenue des Martyrs, F-38042 Grenoble, France}
\affiliation{Univ. Grenoble Alpes, Inst NEEL, F-38042 Grenoble, France}
\email{olivier.bourgeois@neel.cnrs.fr}


\begin{abstract}
	Specific heat of phonon cavities is investigated in order to analyse the effect of phonon confinement on thermodynamic properties. The specific heat of free standing very thin SiN membranes in the low dimensional limit is measured down to very low temperatures (from 6~K to 50~mK). In the whole temperature range, we measured an excess of specific heat orders of magnitude bigger than the typical value observed in amorphous solids. Below 1~K, a cross-over in $c_p$ to a lower power law is seen, and the value of specific heat of thinner membranes becomes larger than that of thicker ones demonstrating a significant contribution coming from the surface. We show that this high value of the specific heat cannot be explained by the sole contribution of 2D phonon modes (Lamb waves). The excess specific heat, being thickness dependent, could come from tunneling two level systems (TLS) that form in low density regions of amorphous solids located on the surfaces. We also show that the specific heat is strongly tuned by the internal stress of the membrane by orders of magnitude, giving unprecedentedly high values, making low stress SiN very efficient for energy storage at very low temperature. 
\end{abstract}


\maketitle

The thermal properties of low-dimensional amorphous material nanowires, membranes etc... are at the heart of essential questions for many subjects such as bolometry, temperature sensing, or simply the understanding of phonon physics at very low temperature in dielectric nanosystems \cite{Giazotto}. Even if the thermal transport in low-dimensional dielectric nanostructures has recently been the subject of numerous although difficult experiments at sub-kelvin temperatures \cite{Schwab,Zen2014,Tavakoli2017,Tavakoli2018}, the specific heat, or the capacity of a body to rise in temperature following a supplied external energy, remains very poorly understood and experimentally almost unexplored \cite{anghel}.

At very low temperatures, two significant physical effects are expected to impact the thermal properties of thin membranes, driving away the specific heat far from the one of a conventional 3D solid material. The first is the effect of the confinement of the phonon modes in nanostructures thinner than the wavelength of phonons. In that kind of free standing membranes, it is expected that new phonon modes will appear having quadratic and not linear dispersion relations, modes whose contribution to the specific heat at low temperature is linear in $T$ \cite{anghel,kuhn,Maasilta,Gusso,fefelov,chavez,Puurtinen2016}. Such modes have been experimentally observed in the thermally relevant hypersonic frequency range \cite{cuffe}. Note that simple arguments would have naively predicted that the specific heat of a confined phonon gas in the low temperature limit should scale with $T^{\delta}$, $\delta$ being the effective dimension of the considered system (for a suspended membrane $\delta =2$ at low enough temperature).  

Secondly, it is well known that the thermal behaviour of amorphous (glassy) materials differs significantly from their crystalline counterparts. For more than forty years, thermal properties of such glassy materials have been explained by the possible presence of dynamic defects acting like tunneling two-level-systems (TLS) \cite{Phillips1972,Anderson1972,Pohl2002}. Below a certain temperature, usually around 1~K, these defects add a sizeable contribution to the specific heat through additional degrees of freedom providing also a linear variation of $c_p$ in temperature. These dynamic defects are expected to scatter phonons, thus acting on the thermal transport of such low dimensional structures \cite{Tavakoli2017,Tavakoli2018}.

Consequently, very particular behaviour of the phonon specific heat is expected at very low temperature in quasi-2D self-suspended amorphous materials. Such systems should thus allow to address the question of the main mechanism governing the thermal properties of confined glassy material when it moves from a standard system described by nuumnuum elastic theory to a system dominated by TLS, or the question of how the internal characteristics of an amorphous material such as stress can influence thermal properties \cite{Wu2011,Southworth,Ftouni}. Apart from the fundamental aspect of differentiating between both theoretical models, since specific heat governs the characteristic time scales of low temperature thermal detectors which often employ suspended ultra-thin membranes, it is essential to solve this controversy by doing direct experiments on them.

Here, we report highly sensitive specific heat measurements of thin membranes in the quasi-2D limit. They were performed on thin silicon nitride suspended structures at sub-Kelvin temperatures using zeptoJoule calorimetry. By comparing results obtained for two different thicknesses (100~nm and 300~nm), it is shown that above 1~K the specific heat is dominated by 3D bulk contributions whereas for $T<1$~K surface contributions dominate. Moreover, even if we observe a close to linear behaviour of the specific heat as a function of the temperature below 1~K, we show that the \textquoteleft standard\textquoteright~elastic continuum model (Lamb modes) cannot explain even the order of magnitude of the measured specific heat. The presence of TLS in the amorphous dielectric is the most probable scenario to explain this anomalously high specific heat. The absolute value of the specific heat measured in our experiment is \emph{up to five orders of magnitude} larger than that measured in a crystalline bulk material, making the used SiN thin membranes the system with the largest specific heat in this range of temperatures.

\begin{figure*}
	\begin{center}
		\includegraphics[width=15cm]{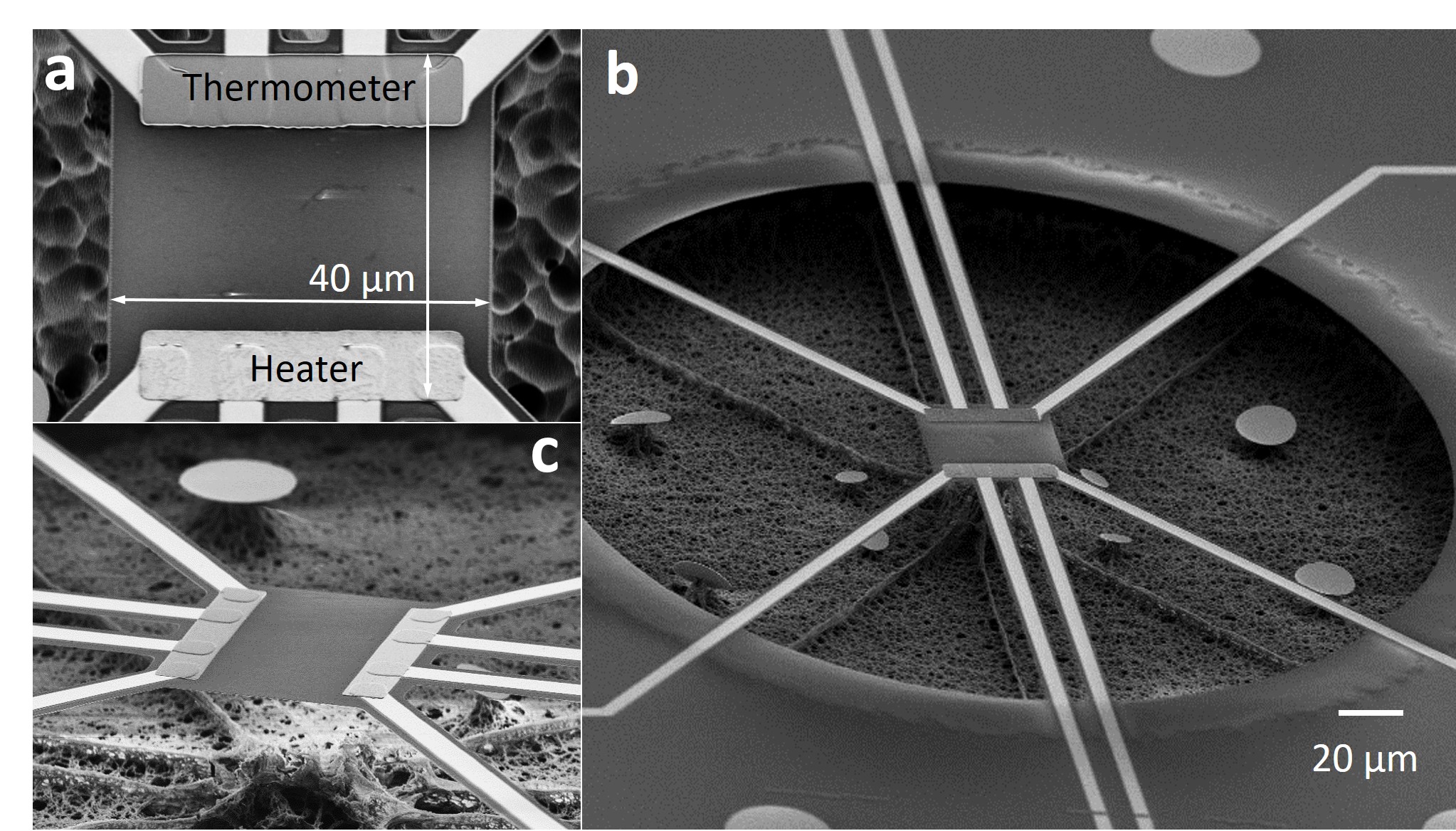}
		\caption{Suspended membrane-based zeptoJoule nanocalorimeter. \textbf{(a)} A top view of the 2D nanocalorimeter made of amorphous SiN that is functionalized with two transducers: a heater and a thermometer. \textbf{(a)}, \textbf{\textbf{(b)}}, and \textbf{(c)} are the angled views of the nanocalorimeter that is suspended by eight suspension beams.}
		\label{fig:1}
	\end{center}
\end{figure*}

Specific heat of suspended thin membranes of silicon nitride has been measured down to very low temperatures using a highly sensitive technique. Various types of silicon nitride films have been used to fabricate the sensors: stoichiometric Si$_3$N$_4$ having a high internal tensile stress
 of 0.95~GPa, and two non-stoichiometric silicon nitrides, one low stress (0.2~GPa) and one super low stress ($<0.1$~GPa). We used 100~nm and 300~nm thick SiN films commercially available deposited by low pressure chemical vapor process on top of a Si substrate. The thicknesses have been chosen to be of the same order of magnitude as the dominant phonon wavelength to ensure that only the 2D phonon modes are occupied in the membrane. Here we define the dominant phonon wavelength $\lambda_{dom}$ to correspond to the wavelength at which the Planck distribution (spectral energy density per volume) for bosons is maximum \cite{Bourgeois2016}. For 3D phonons, the dominant phonon wavelength is given by: 

\begin{equation}
\lambda_{dom}= \frac{h v_{s,i}}{2.82 k_B T}
\label{lambda}
\end{equation}

with $v_{s,i}$ the velocity of sound for mode $i$, $h$ is the Planck constant and $k_B$ the Boltzmann constant; for SiN $v_s=10000$~m~s$^{-1}$, $\lambda_{dom}=100$~nm around 1~K. By fabricating 100~nm thick membranes, the 2D phonon regime for thermal properties should be accessible at the lowest temperature of the experiment (below 0.1~K).

The thermal sensor is made of a self-supported SiN membrane of size 40~$\mu$m by 40~$\mu$m defining the dimensions of the quasi-2D phonon cavity suspended by eight arms (6~$\mu$m wide and 100~$\mu$m long). The long suspending arms lead to a very good thermal insulation of the membranes from the Si frames \cite{Tavakoli2018}. These arms serve also as mechanical support for the electrical leads of the transducers (see Fig.~\ref{fig:1}). A copper heater and a niobium nitride thermometer (NbN) are structured by laser lithography on the SiN membrane as shown in Fig.~\ref{fig:1}~a, the electrical leads are made with superconducting materials (NbTi). The temperature coefficient of resistance of the NbN calculated through $\alpha=\frac{1}{R} \frac{dR}{dT}$ can be higher than $-1$~K$^{-1}$, a particularly high value for thin film technology \cite{Tuyen2019}.

 \begin{figure*}
 	\begin{center}
 		\includegraphics[width=12cm]{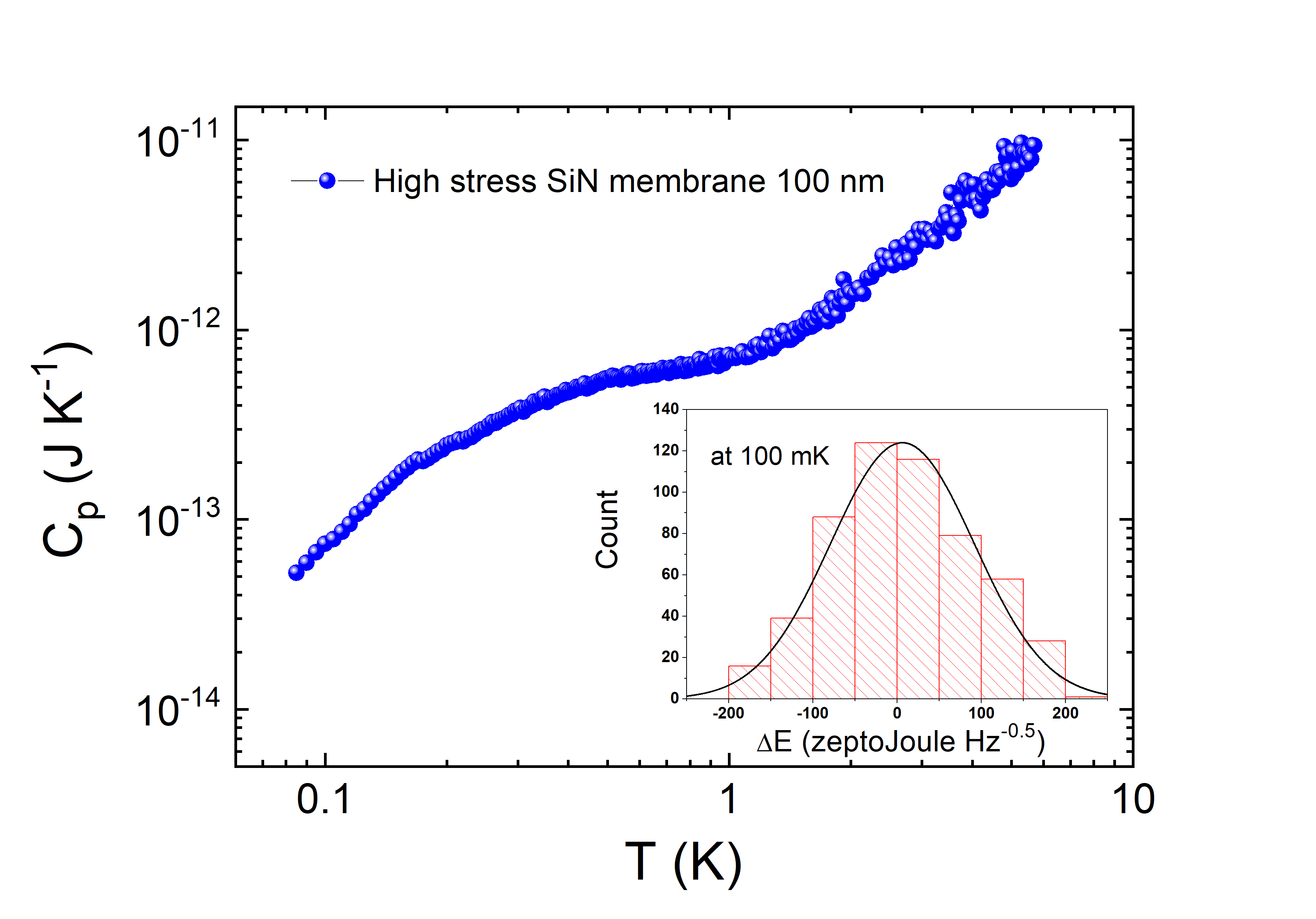}
 		\caption{Heat capacity of a 100~nm thin membrane made out of high-stress Si$_3$N$_4$. The measurement of the heat capacity is shown over a temperature range of 0.07~K to 7~K. The noise has been measured at 100~mK as shown in inset. The full width at half maximum shows that the measurement is done with a sensitivity in energy of $\pm$~100 zeptoJoule.}
 		\label{Cpraw}
 	\end{center}
 \end{figure*}

An example of a measurement of heat capacity is presented in Fig.~\ref{Cpraw}. It is measured using an ac-calorimetry technique, a dynamic method based on modulating the temperature being efficient and highly sensitive especially at low temperatures. In recent papers \cite{Bourgeois2005,Souche2013,Poran2014,Poran2017}, we have demonstrated on these sensors a sensitivity for specific heat measurement of $2 \times 10^{-16}$~J~K$^{-1}$~$\sqrt{ \rm{Hz}}^{-1}$, for an oscillation of temperature of 3~mK, giving an unprecedented sensitivity in energy of $\pm 10^{-19}$~J~$\sqrt{ \rm{Hz}}^{-1}$, i.e. $\pm$~100~zeptoJoule per $\sqrt{ \rm{Hz}}^{-1}$ at 100~mK.
The sample holder is installed in a vacuum chamber ($<2 \times 10^{-6}$~mbar before cryopumping) on a stage regulated with a precision better than 20~$\mu$K at a temperature of 100~mK measured using a calibrated Speer carbon thermometer \cite{Tuyen2019}.
This set-up allows the measurement of the specific heat of 2D phonon cavities from below 100~mK to above 7~K as it can be seen in Fig.~\ref{Cpraw}.

\begin{figure*}
	\begin{center}
		\includegraphics[width=20cm]{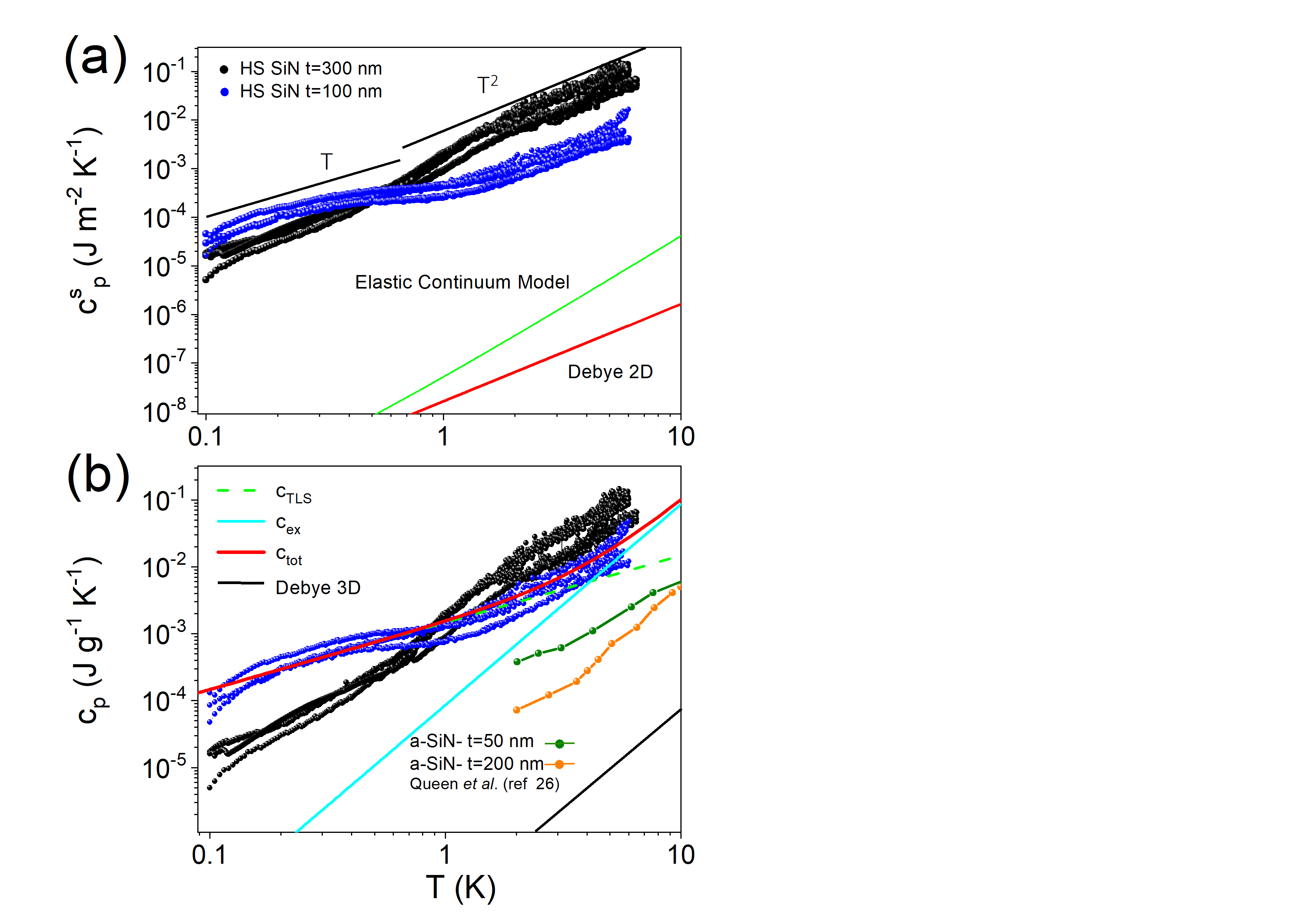}
		\caption{The specific heat of thin membranes made of high-stress Si$_3$N$_4$. (\textbf{a}) The heat capacity normalized by the surfaces area of the membranes, in comparison to the predicted specific heat for 2D phonon cavity for two sets of membranes of  thicknesses 100~nm and 300~nm. The green lines are the prediction of elastic continuum theory, and the red line is the 2D Debye model. The two black straight lines are guides to the eyes for the temperature power laws $T$ and $T^2$ (\textbf{b}) The heat capacity normalized by volume, same data as in (a). The blue line is the excess of specific heat $C_{ex}$ from Eq.~\ref{Cex}, the green dashed line is the specific heat of TLSs $C_{TLS} \propto T$, and the red line is the total estimated specific heat from the sum of Eq.~\ref{CTLS} and Eq.~\ref{Cex} for an amorphous solid from which the TLS density is extracted. The black line is the predicted specific heat from the de Debye 3D model. The green and orange dots are the specific heat of amorphous SiN of thickness 50~nm and 200~nm respectively, measured by Queen \textit{et al} \cite{Queen2009} plotted for comparison.}
		\label{Cpexp}
	\end{center}
\end{figure*}

Specific heat measurements done on two sets of membranes of high-stress SiN of thicknesses 100~nm and 300~nm are presented in Fig.~\ref{Cpexp} in different units: 1-in units normalized by the surface area J~m$^{-2}$~K$^{-1}$ highlighting the surface effects in panel (a), and 2-in the regular units J~g$^{-1}$~K$^{-1}$ for volumetric specific heat highlighting bulk effects in panel (b). As expected, the specific heat (surface or volumetric) decreases when lowering the temperature as less and less degrees of freedom are excited. The total heat capacity signal is dominated by the SiN membrane contribution, as the heat capacities of the copper heater and the NbN thermometer are at least two orders of magnitude smaller over the whole temperature range of the experiments and will be neglected in the following discussion.

If we focus first on the low temperature part, one clearly sees that below 1~K, both thin and thick membranes have approximately the same surface specific heat (see Fig.~\ref{Cpexp}a), whereas in the high temperature limit (above 1~K), it's the specific heat per unit volume which is almost identical for both types of samples (see Fig.~\ref{Cpexp}b). This indicates that in the low temperature range, the specific heat of the samples seem to be governed by the surfaces, whereas at higher temperatures, the specific heat is dominated by volumetric effects. One basis for interpretation resides in comparing the thickness of the samples with the dominant wavelength of the phonons. Indeed, a possible change between a 3D volume effect and a 2D surface effect arises when the phonon dominant wavelength becomes bigger than the membrane thickness.
As mentioned earlier, by using Eq.~\ref{lambda}, we obtained that $\lambda_{dom}$ is on the order of few hundreds of nanometer around 1~K. One can conclude that at temperatures lower than that the phonon specific heat is dominated by the lowest 2D phonon modes, as their wavelength is much larger than the thickness of the sample. As the temperature increases, $\lambda_{dom}$ decreases and one recovers the standard three dimensional behaviour for the specific heat. Considering these two limits, we can infer that we expect a 2D-3D dimensional cross-over in the specific heat in this system, the cross-over temperature being below 1~K. Then, the following questions need to be addressed: can we attribute the change of specific heat with temperature to a phonon dimensionality effect? does it appear with the correct order of magnitude? In the following, we will estimate the Debye 3D, Debye 2D and elastic continuum model (ECM) of specific heat and compare the different orders of magnitude obtained to sort out the various physical mechanisms at play.

The Debye 3D model of specific heat is characterized by its temperature variation following a cubic power law $C \propto T^3$ as given by the formula \cite{cleland}:
\begin{equation}
c_{Debye 3D}= \frac{2 \pi^2 k_B^4}{\rho 15 \hbar^3} \left( \frac{2}{v_T^3} +\frac{1}{v_L^3} \right)  T^3
\label{debye3D}
\end{equation}

where $\rho$ is the mass density in g~m$^{-3}$, $\hbar$ is the reduced Planck constant, and $k_B$ is the Boltzmann constant, $v_T$ and  $v_L$ the transversal and longitudinal speeds of sound respectively 6200~m~s$^{-1}$ and 10300~m~s$^{-1}$ for silicon nitride \cite{kuhn}; the specific heat $c_{Debye 3D}$ is in J~g$^{-1}$~K$^{-1}$ . This model has demonstrated over the years its great ability to fit $c_p$ experimental values of crystalline electrically insulating materials \cite{Rosenberg,Gopal}. It can be adapted to 2D systems by integrating over the wavevectors parallel to the membrane surface in a planar geometry (keeping the regular longitudinal and transverse 3D bulk phonon modes):

\begin{equation}
c_{Debye 2D}= \frac{3 \zeta (3) k_B^3}{\pi \hbar^2} \left( \frac{2}{v_T^2} +\frac{1}{v_L^2} \right) T^2
\label{debye2D}
\end{equation}

where $\zeta$ is the Riemann zeta function. This area dependent specific heat in J~m$^{-2}$~K$^{-1}$ does not depend on the thickness of the 2D membrane and has a lower temperature exponent as compared to the Debye 3D model, the power law being given by the dimension of the system. This Debye 2D model is compared to the experimental surface specific heat as measured on SiN membranes in Fig.~\ref{Cpexp}a. Even if the Debye 2D specific heat is larger than the Debye 3D one, significant discrepancies still exist. Indeed, the measured absolute value of $c_p$ is in strong disagreement with the calculated ones, being few orders of magnitude bigger.
 
A refinement of the model seems necessary. This is done by taking into account the modified phonon modes present in thin membranes. In the following, the specific heat of thin free-standing membranes at low temperature is obtained through the calculation of phonon modes of a phonon cavity using the elastic continuum model  \cite{kuhn,Gusso,fefelov,chavez,Puurtinen2016}. In this model, some bulk polarizations will couple together at the surfaces generating new set of vibrational modes: dilatational waves (or symmetric Lamb waves), flexural waves (or asymmetric Lamb waves) and horizontal shear waves \cite{Karl,fefelov,chavez}. The modified 2D dispersion relations of these modes have important consequences on the low temperature specific heat. Again, by integrating over the wavevector parallel to the membrane surface $q_{\mid \mid}$, the total surface specific heat can be written as:

\begin{equation}
c_{ECM}= \frac{A}{k_BT^2 2\pi} \sum_\sigma \sum _{m=0}^{\infty} \int_0^\infty dq_{\mid \mid} \frac{q_{\mid \mid} (\hbar \omega_{m,\sigma})^2 \exp(\beta \hbar \omega_{m,\sigma})}{\left[\exp(\beta \hbar \omega_{m,\sigma})-1\right]^2}
\end{equation}

where $\beta=(k_BT)^{-1}$, $\sigma$ represents the different modes and $\sum_m$ is the summation over all the branches of a mode.

According to the calculations at low enough temperature, for membranes much thinner than $\lambda_{dom}$, the contribution of the flexural waves becomes dominant. In contrast to the Debye 2D model, the elastic continuum theory then predicts a linear variation of specific heat (not quadratic) with temperature at very low temperature \cite{Gusso,kuhn,chavez}. This linear dependence comes from the quadratic dispersion relation of the dominant flexural modes. The temperature at which this 3D-2D cross-over appears depends on the thickness of the phonon cavity via the dependence on temperature of the dominant phonon wavelength. The surface specific heat, as evaluated by ECM, is shown in Fig.~\ref{Cpexp}a only above 0.5~K due to its very low value. The measured $c_p(T)$ and the calculated one by ECM have very dissimilar absolute values showing that the elastic continuum model do not give a good account of what has been measured by few orders of magnitude.

All we have considered so far are in principle valid only for \emph{crystalline} systems. As SiN thin films are amorphous and disordered, some specific thermal properties may be expected. One of the most important properties of non-crystalline materials is that they contain a large amount of two-level systems. Two level systems are atoms or groups of atoms that have two metastable configurations in their energy landscape. At low enough temperature these atoms or groups of atoms can switch between these two states only by quantum tunneling. An important ingredient when considering the physics of these TLS is that a uniform and broad spectral distribution for time constants or equivalently barrier heights is assumed. The exact origin of these two level systems is still heavily debated, though they can be seen as extra degrees of freedom in the solid \cite{leggett1991,Wu2011,leggett2013,Hellman2006,zink2004,Hellman2013,Hellman2015,Lubchenko}. At low temperature, phonons can  interact strongly with the TLSs, having not only consequences on the specific heat but also on the phonon thermal transport \cite{Pohl2002,Tavakoli2017,Tavakoli2018}. The contribution of the TLSs to the specific heat is given by \cite{Anderson1972,Wu2011,Hellman2015}:

\begin{equation}
c_{TLS}=\frac{\pi^2}{6}\frac{n_0}{\rho}k_{B}^{2}T=\alpha T
\label{CTLS}
\end{equation}

with $\rho$ the mass density in g~m$^{-3}$ of the material and $n_0$ the density of TLS; $\alpha$ is in J~g$^{-1}$~K$^{-2}$. $c_{TLS}$ is the specific heat linked to low energy excitation specific to glassy materials, it will dominate the $c_p$ signal at low temperature. Another contribution to specific heat in amorphous solids can come from extra non-propagating modes; this is usually modeled by an excess specific heat having a cubic power law \cite{Hellman2015}:

\begin{equation}
c_{ex}=\gamma\,{T^3}
\label{Cex}
\end{equation}

$\gamma$ being a factor determined experimentally. For most disordered materials, $\gamma$ lies in the range 1 to 10$\times 10^{-7}$ J~g$^{-1}$~K$^{-4}$. This excess specific heat has been observed in most disordered materials \cite{Pohl2002,Hellman2015}. Non-propagating modes and TLS are both associated with low density regions specific to disordered materials where voids or floppy bonds are present. The total specific heat of amorphous materials will be given by the relation $c_{tot} =c_{TLS}+c_{ex}=\alpha T+ \gamma T^3 $.

We can now compare the contribution of these two different mechanisms in disordered solids to the measured specific heat. In Fig.~\ref{Cpexp}b, we have reported our experimental data for the volumetric specific heat and plotted the theoretical contributions to $c_p$, the excess specific heat and the specific heat due to two level systems, as obtained from Eq.~\ref{CTLS} and Eq.~\ref{Cex}; the prefactors $\gamma$ in Eq.~\ref{Cex} and the density of two level system $n_0$ in Eq.~\ref{CTLS} being the adjustable parameters. We have also plotted data taken by other groups on 50~nm and 200~nm thick SiN membranes \cite{Pohl2002,Queen2009}. By comparing the different experimental data, we can first notice that the specific heat we observe is larger by more than two orders of magnitude than the one measured on a-SiN membranes . 

In order to reproduce the experimental data using the TLS model, we have to consider the specific heat due to localized soft modes ($c_{ex}$, Eq.~\ref{Cex}) as well as the specific heat due to two level systems ($c_{TLS}$, Eq.~\ref{CTLS}), taking for the adjustable parameters $\gamma \approx 10^{-4}$ J~g$^{-1}$~K$^{-4}$ and $n_0\approx 1.5\times 10^{49}$ J$^{-1}$~m$^{-3}$. Using these parameters, we can reproduce \emph{quantitatively} the experimental data as shown in Fig.~\ref{Cpexp}b with the red line fit. The membrane specific heat is dominated over the whole temperature range of the experiment by the amorphous nature of the SiN: above 1~K, the specific heat is interpreted to be dominated by localized soft modes, whereas in the low temperature range, the specific heat is almost entirely due to the presence of two level systems. The density of TLS in the 100~nm thick membrane is $n_0\approx 1.5\times 10^{49}$ J$^{-1}$~m$^{-3}$  a relatively high value as for 300~nm $n_0\approx 1.5\times 10^{48}$ J$^{-1}$~m$^{-3}$, lying slightly above the measured range in amorphous silicon $n_0\approx  10^{44}$ J$^{-1}$~m$^{-3}$ to $n_0\approx 10^{48}$ J$^{-1}$~m$^{-3}$ \cite{Hellman2015}.

To confirm this scenario of TLS dominated $c_p$, it is interesting to try to control the density of the two level systems in the material; this can be done by changing the mass density of the material by the following mechanism: when the composition of SiN is changed this affects the internal stress along with the mass density. Since the TLS are thought to result from dangling bonds arising from voids or surfaces, varying the mass density and stress should affect the TLS density and with it the specific heat. We consequently expect a higher TLS density in lower mass density materials having lower mechanical stress. Therefore, we decided to reproduce the same $c_p$ experiments on high-stress (HS), low-stress (LS) and super low-stress (SLS) amorphous silicon nitride membranes to probe the effect of various TLS density on the specific heat.

\begin{figure*}
	\begin{center}
			\includegraphics[width=12cm]{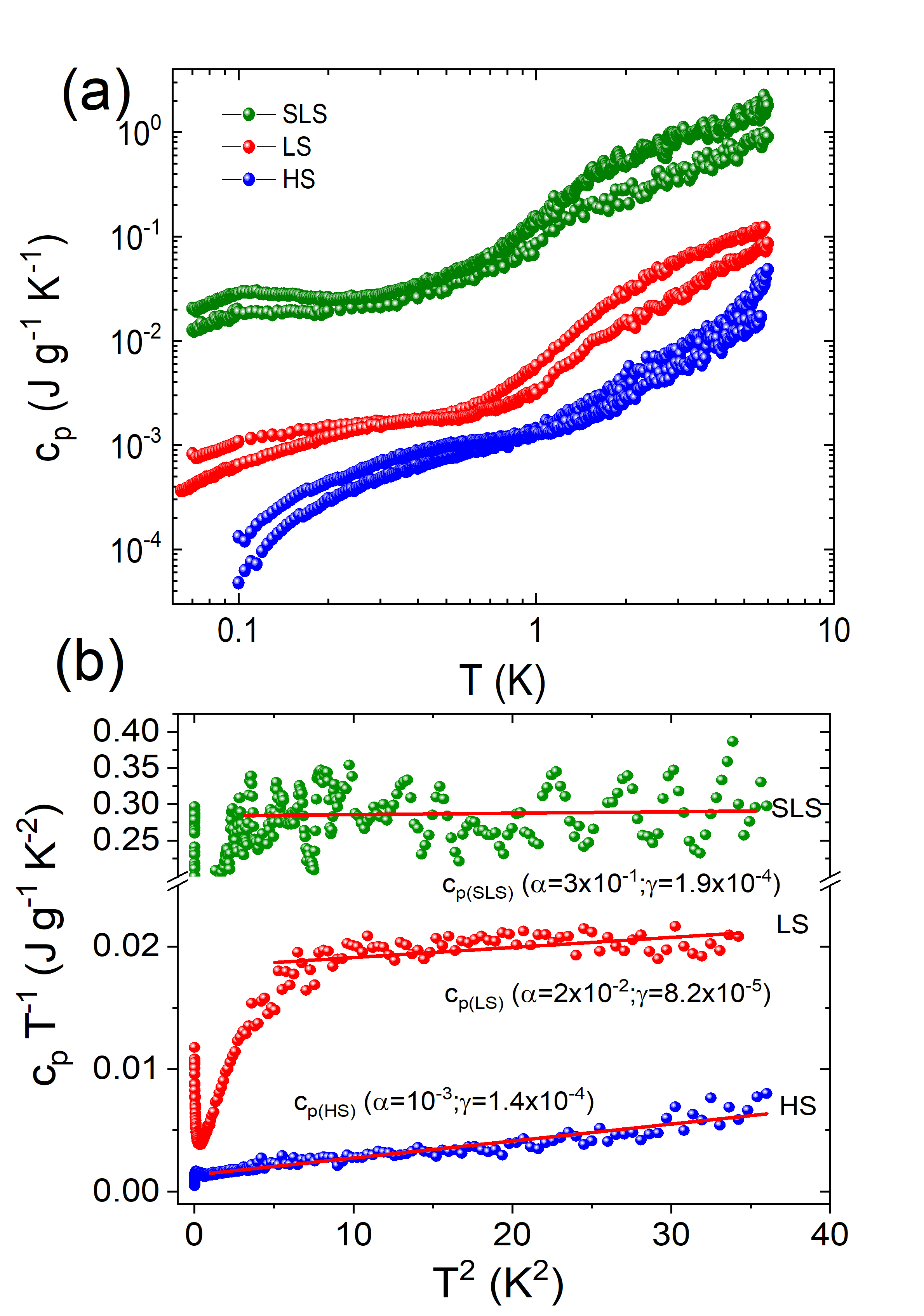}
		\caption{The temperature dependence of the specific heat for different internal stress of SiN$_x$. The samples have all the same geometry with a thickness of 100~nm. The specific heat measurements have been done for two sets of membranes of each kind of SiN: blue, red and green dots are respectively the specific heat of high-stress (HS), low-stress (LS) and super low-stress (SLS) amorphous silicon nitride. In (a) the specific heat vs $T$ is plotted in a log-log scale and in (b)  $c_p/T$ versus $T^2$  to highlight the $\alpha$ (intercept with $y$ axis) and $\gamma$ (the slope of the linear fit) values for each internal stress. The unit of $\alpha$ is J~g$^{-1}$~K$^{-2}$  and for $\gamma$ J~g$^{-1}$~K$^{-4}$.}
		\label{Cstress}
	\end{center}
\end{figure*}

The temperature dependence of the specific heat of the different 100~nm thick SiN membranes is depicted in Fig.~\ref{Cstress}. Using Eq.~\ref{CTLS} we can extract from our measurements an order of magnitude for the concentration of TLS in our membranes, summarized in the Table~\ref{tableSiN}. This table displays the composition, the mass density, the TLS density and the stress for each composition as given by the supplier \cite{simat}. The composition and the mass densities were determined by Rutherford Backscattering Spectrometry (RBS).

\begin{table*}
\begin{center}
\begin{tabular}{l c  c  c }

\hline
\hline\noalign{\smallskip}
  SiN$_x$ & HS & LS & SLS \\
  
  composition & Si$_{43}$N$_{57}$ & Si$_{46}$N$_{54}$ & Si$_{48}$N$_{52}$ \\
    
$\rho$ (g~cm$^{-3}$)&  3.2 & 2.9 & 2.4$\pm$0.2\\

$n_0 $ (J$^{-1}$~m$^{-3}$)  & 1.5$\times 10^{49}$  & 3.6 $\times 10^{49}$ & 3.3 $\times 10^{50}$\\

stress (GPa) & 0.85 & 0.2 & $<$0.1\\

\hline
\hline
	\end{tabular}
		\caption{Parameters for high-stress (HS), low-stress (LS), and super low-stress (SLS)  SiN$_x$. $\rho$ is the mass density of material measured by Rutherford Backscattering Spectrometry, $n_0 $ is the TLS density of states as extracted from the measurements using Eq.~\ref{CTLS}, and the stress for each composition, as given by the supplier \cite{simat}.}
\label{tableSiN}
\end{center}
\end{table*}

The main observation from Fig.~\ref{Cstress} is that by decreasing the stress and the mass density between HS and SLS SiN, the specific heat is increased by two orders of magnitude. These values are surprisingly high in comparison to what has been measured until now on amorphous dielectric materials \cite{Pohl,Wu2011,Hellman2015}. We suppose that such high value of the specific heat can be attributed to the presence of two level systems and localized modes in amorphous SiN. One remarkable feature appears when inspecting Table~\ref{tableSiN}: it seems that there is a correlation between the density of TLS and the mass density. Indeed, there is a factor of twenty between the density of TLS in HS and SLS a-SiN$_x$. In between these two samples, the mass density varies by more than $30\%$. We guess that this variation of mass density is due to the appearance of low density regions in the amorphous SiN located on the surface; it is precisely these low density regions which are favourable for the formation of TLS; this could explain the difference by orders of magnitudes of specific heat between SLS and HS SiN through a much higher TLS density. Such a behaviour has never been reported before; more investigations will be required to elucidate its true origin.

In conclusion, we have measured specific heat of thin SiN membranes from 6~K down to very low temperature (50~mK) to investigate the heat capacity of 2D phonon cavities. In the whole temperature range, the specific heat does not exhibit a single power law as a function of temperature but a cross-over between two different regimes around 1~K. As it has been suggested first by Gusso \textit{et al.} \cite{Gusso}, this cross-over could have been correlated to a 3D-2D transition when the phonon wavelength becomes of the order of the thickness of the membrane. However, first, the position of the cross-over in temperature is not in agreement with the ECM model, it is at too high temperature and second the absolute value of the measured specific heat is much larger than that predicted from our ECM analysis in terms of surface specific heat. This is ruling out that, in SiN membranes, the contribution of the Lamb modes to $c_p$ can explain such very high values of specific heat. Rather, we have demonstrated that the exceptionally high magnitude of the specific heat likely results from the presence of a large density of two level systems, as expected for the amorphous disordered systems as the ones used in this work (SiN membranes). Our modeling indicates that above 1~K the specific heat is dominated by an excess bulk-like contribution, possibly due to non-propagating modes, and by two-level systems at lower temperatures. Using different SiN compositions, we have also shown that by changing the density of two-level systems through the mass density of the SiN material, unprecedentedly high values of specific heat can be reached \cite{Pohl,Wu2011,Ramos}. The formation of TLS is enhanced by to the presence of low-density regions in the SiN, and the presence of large surfaces due to the membrane geometry of the samples. These results will have a significant impact on the performances of low temperature sensors.

The authors thank the technical support provided by the Nanofab, P\^{o}le Capteur, P\^{o}le \'{e}lectronique, P\^{o}le de cryog\'{e}nie facilities in Institut N\'{e}el. We thank Cyril Bachelet and Claire Marrache-Kikuchi from Irene Joliot-Curie Lab-CNRS for their help in the RBS measurements and Boris Brisuda for proof reading. The research leading to these results has received funding from the European Union's Horizon 2020 Research and Innovation Programme, under grant agreement No.~824109, the European Microkelvin Platform (EMP), the EU project MERGING grant No.~309150, ERC CoG grant ULT-NEMS No.~647917, the authors also acknowledges the financial support from the ANR project QNM Grant No.~040401. T.P. and I.M. acknowledge the financial support from the Academy of Finland project number 341823.

\end{document}